\documentclass[conference]{IEEEtran}
\usepackage{amsmath,epsfig}
\usepackage{algorithm}
\usepackage{bm}
\usepackage{tikz}
\usepackage{graphicx}
\usetikzlibrary{positioning}
\usetikzlibrary{shapes.multipart}
\usepackage{pgfplots}
\usepackage[usenames,dvipsnames]{pstricks}
\usepackage{epsfig}
\usepackage{pst-grad} 
\usepackage{pst-plot} 
\usepackage{subfigure}
\usepackage[capitalize]{cleveref}

\usepackage{subfigure}
\usepackage[T1]{fontenc}
\usepackage{amsmath}
\usepackage{graphics}
\include{psfig}
\usepackage{epsfig}
\usepackage{array}
\usepackage{psfrag}
\usepackage{amssymb}
\usepackage{color}
\usepackage{pstricks,pst-node,pst-text,pst-3d,pst-plot}
\usepackage{cite}

\usepackage{ulem}
\definecolor{dgreen}{rgb}{0, 0.8, 0.1}

\def\bg{{\bf g}}

\def\bR{{\bf R}}

\def\bE{{\bf E}}
\def\bd{{\bf d}}

\def\ba{{\bf a}}

\def\bs{{\bf s}}

\begin{document}

\title{CDM Based Virtual FMCW MIMO Radar Imaging at 79 GHz}

\author{\IEEEauthorblockN{Shahrokh Hamidi}
\IEEEauthorblockA{Department of Electrical and Computer Engineering\\
University of Waterloo\\
Waterloo, Ontario, Canada\\
Email: shahrokh.hamidi@uwaterloo.ca}
\and
\IEEEauthorblockN{Safieddin Safavi Naeini}
\IEEEauthorblockA{Department of Electrical and Computer Engineering\\
University of Waterloo\\
Waterloo, Ontario, Canada\\
Email: safavi@uwaterloo.ca}
}

\maketitle

\begin{tikzpicture}[remember picture, overlay]
      \node[font=\small] at ([yshift=-1cm]current page.north)  {This paper has been accepted for publication in the IEEE Canadian Conference on Electrical and Computer Engineering, 2021. \copyright IEEE};
\end{tikzpicture}

\begin{abstract}
In this paper, we will be using a $\rm 79 \;GHz$ Multiple Input Multiple Output (MIMO) Frequency Modulated Continuous Wave (FMCW) radar and apply the Code Division Multiplexing (CDM) method to increase the number of elements virtually. This, in turn, enhances the angular resolution of the radar. The major contribution of our work comes from the fact that by exploiting the CDM  method we will make it possible for multiple FMCW radars to operate in close proximity of each other. In fact, using coded signals the effect of interference can be alleviated considerably.

\end{abstract}


\section{Introduction}
Multiple Input Multiple Output (MIMO) Frequency Modulated Continuous Wave (FMCW) radars operating at $\rm 79 \;GHz$ are compact, light, and cost effective devices with low peak-to-average power ratio. They have applications in different areas such as automotive industry and Unmanned Arial Vehicle (UAV) based radar imaging. In order to keep the structure small and simple, these radars come with a small number of transmitters and receivers.
The number of elements can be virtually increased using techniques such as Time Division Multiplexing (TDM), Frequency Division Multiplexing (FDM), or Code Division Multiplexing (CDM), and as a result, higher angular resolution can be achieved. Both the TDM and the FDM-based virtual FMCW MIMO radar imaging systems have been reported in the literature \cite{Shahrokh, 24GHzTDM, 77GHzTDM, TDMHindawi,TDMcalibration,TDMmotioncompensation,FDMMIMO}. However, to the best of our knowledge, the CDM-based virtual FMCW MIMO radar has not received any attention

Compared to the FDM-based virtual FMCW MIMO radar, the TDM-based virtual FMCW MIMO radar has simpler structure and higher resolution in the range direction. However, since in the TDM-based virtual FMCW MIMO radar transmitters transmit at different time slots, therefore, the signal to noise ratio (SNR) is lower compared to the FDM-based virtual  FMCW MIMO radar in which all the transmitters transmit simultaneously.

The lower resolution for the FDM-based virtual FMCW  MIMO radar comes from the fact that in the FDM-based virtual FMCW MIMO radar the entire bandwidth is divided into several pieces to create the orthogonal beams in space. As a result, each transmitter uses only a portion of the entire bandwidth and the range resolution decreases.

The CDM-based virtual FMCW MIMO radar, similar to the TDM-based virtual FMCW MIMO radar, has a simple structure and since each transmitter utilizes the full bandwidth of the system, hence, the maximum range resolution can be achieved. Yet, since all the transmitters operate simultaneously, therefore, the SNR will be higher than that of the TDM-based virtual FMCW MIMO radars.

Another great advantage of the CDM-based virtual FMCW MIMO radar stems from the fact that by using orthogonal codes for different radars operating in the same neighborhood, the interference among different users can be avoided. Interference avoidance plays an important role in areas such as automotive applications.

In this paper, we address the CDM-based virtual FMCW MIMO radar imaging. We present the full formulation of the imaging procedure based on the Multiple Signal Classification (MUSIC) algorithm \cite{Music, Music_stoica, stoica}.

We also propose a simple method for the system calibration at the baseband level. We then use experimental data, gathered from a $\rm 79 \; GHz$ FMCW MIMO radar which operates based on the CDM method, and present the result. In order to create orthogonal signals based on the CDM method, we use Walsh codes \cite{Walsh}.

The paper is organized as follows.
In Section \ref{system model}, we present the problem formulation. We address the system calibration at the baseband level in Section \ref{calibration}.
We will then apply the MUSIC technique for image reconstruction in Section \ref{image reconstruction}. Finally, we have dedicated Section \ref{result} to the results of applying the algorithm presented in Section \ref{image reconstruction} to the experimental data gathered from a $\rm 79 \; GHz$ CDM-based virtual FMCW MIMO radar followed by the concluding remarks.

\section{System Model}\label{system model}
The signal transmitted by the $m^{\rm th}$ transmitter and received at the location of the $n^{\rm th}$ receiver after hitting the $l^{\rm th}$ target, is a chirp signal modeled as follows,
\begin{eqnarray}
\label{initial_signal}
s^{(l)}_{mn}(t) = \sigma_l \; e^{\displaystyle  j \left(2 \pi f_c (t - \tau_{mnl}) +  \pi \beta {(t - \tau_{mnl})}^2\right)},
\end{eqnarray}
where $f_c$ is the carrier frequency and $\sigma_l$ is the radar cross section for the $l^{\rm th}$ target. The parameter $\beta$ is given as $b/T$, where $b$ and $T$ stand for the bandwidth and the chirp time, respectively. Finally, $\tau_{mnl}$ is the time delay for the signal to travel from the $m^{\rm th}$ transmitter and be received at the location of the $n^{\rm th}$ receiver after bouncing off the $l^{\rm th}$ target.

\begin{figure}
\centering
\includegraphics[height=8cm,width=8cm]{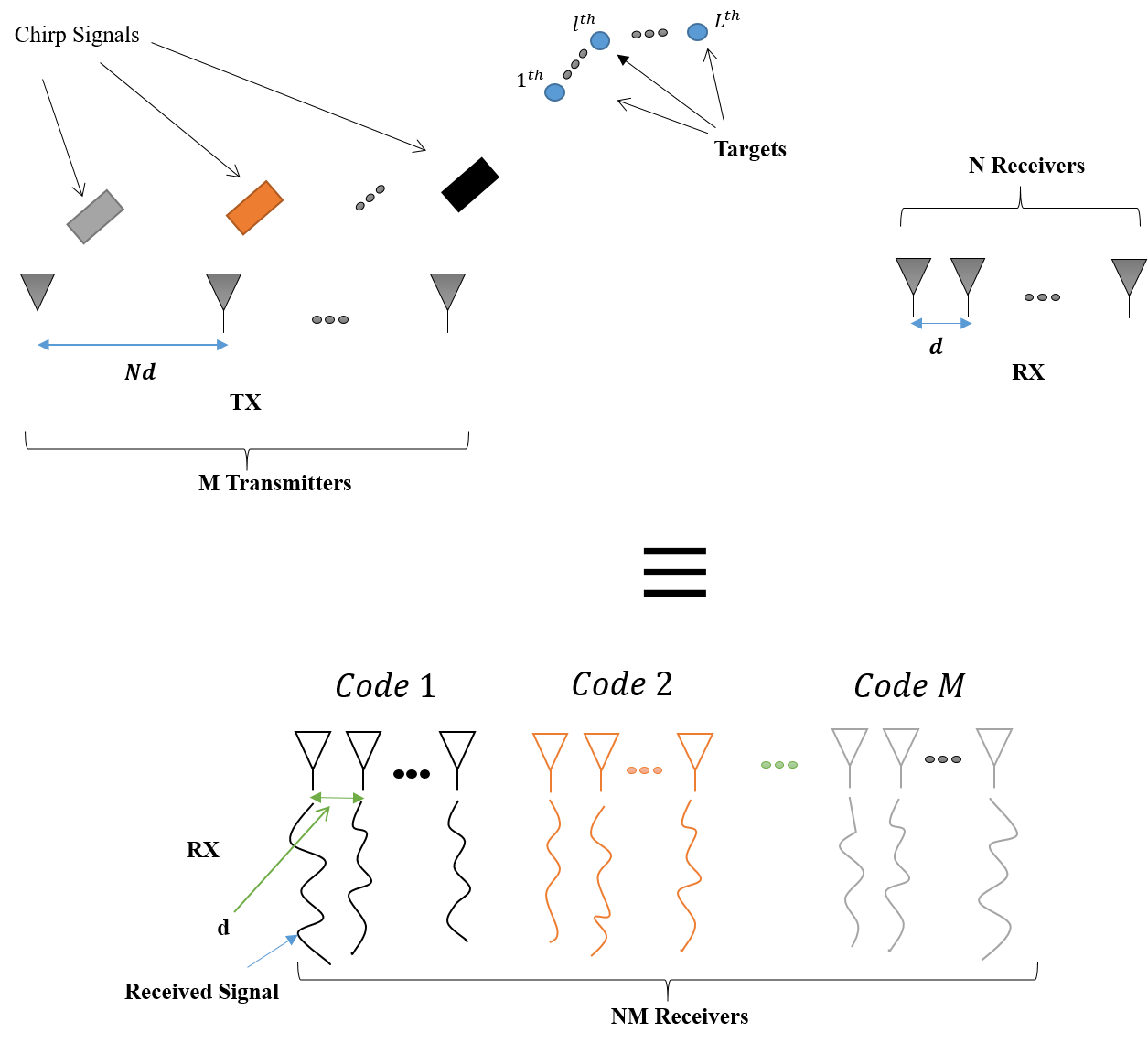}
\caption{A schematic of a virtual MIMO radar based on the CDM method.
\label{fig:CDM_geometry}}
\end{figure}

Fig.~\ref{fig:CDM_geometry} shows the idea behind virtual MIMO radar based on the CDM method.
In the case of the CDM-based MIMO radar, each chirp is transmitted with a specific phase. In this paper, we only consider two values for the phase, $0$ and $180$ degree. Upon applying orthogonal codes, we can then create orthogonal signals in space. The up-chirp signal for the case of the CDM-based virtual FMCW MIMO radar is given as
\begin{align}
\label{initial_signal_CDM}
& {s}^{(l)}_{mni}(t) = \nonumber  \\
 & \sigma_l \; e^{\displaystyle  j \left(2 \pi f_c (t - \tau_{mnl}) +  \pi \beta {(t - \tau_{mnl})}^2 +  \varphi_i(m)\right)}.
\end{align}
In (\ref{initial_signal_CDM}), $\varphi_i(m) \in \{0, \pi\}$ where $i \in \{ 1,2, \cdots, N_c \}$ and $N_c$ is the length of the code. The dependency of $\varphi_i$ on $m$ is the indication of the fact that for each transmitter a unique code is used.

\begin{figure}
\centering
\includegraphics[height=4cm,width=8cm]{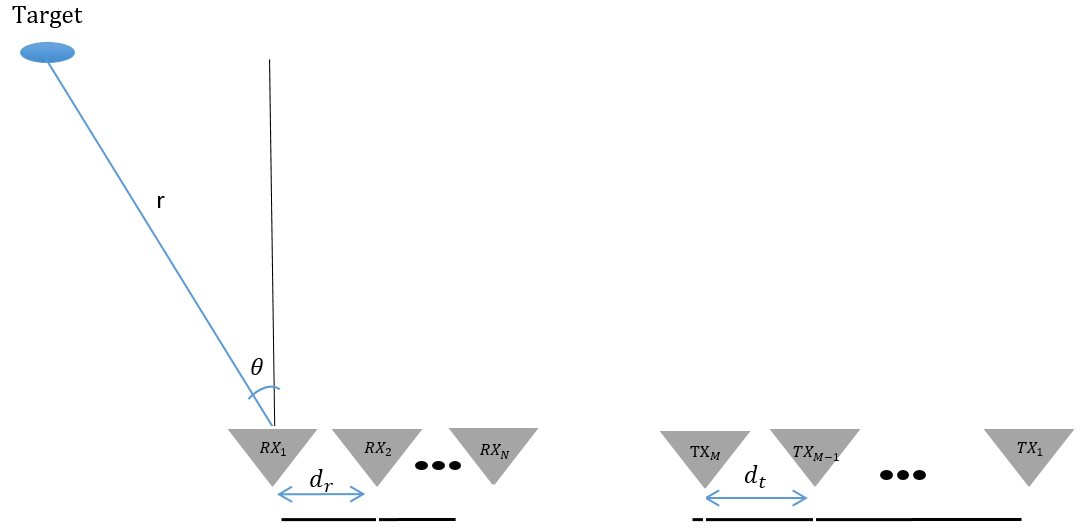}
\caption{Set-up geometry for a MIMO array with M transmit and N receive antennas.
\label{fig:set_up}}
\end{figure}

After the signal is received at the receiver, it will be mixed with a copy of the transmitted signal and the result would be the beat signal. Based on (\ref{initial_signal_CDM}), the beat signal for the signal transmitted by the $m^{\rm th}$ transmitter and received at the location of the $n^{\rm th}$ receiver is then expressed as,
\begin{align}
\label{beat_signal_CDM}
& s^{(l)}_{mni}(t) =  \nonumber \\
& \sigma_l \; e^{\displaystyle  - j \left( 2 \pi f_c \tau_{mnl} +  2 \pi \beta t \tau_{mnl} -  \pi \beta \tau_{mnl}^2 -  \varphi_i(m)\right)}.
\end{align}
Fig.~\ref{fig:set_up} illustrates a schematic of our set-up. In Fig.~\ref{fig:set_up}, we have M transmit and N receive antennas. Parameters $r$ and $\theta$ are the range and incidence angle of a target in front of the array, respectively. Finally, $d_r$ is the element spacing between receive antennas and $d_t$ is the element spacing between transmit antennas.
We place receiver number one at the origin and it will be our reference point as has been shown in Fig.~\ref{fig:set_up}. Our processing will be based on this choice for the reference point. Based on this choice for the reference point, the time delay $\tau_{mnl}$ can be written as $\tau_{mnl} = \frac{2}{c}(r_l + 0.5 d_{mn} sin(\theta_l))$, in which the $MN \times 1$ vector $\bd$ is expressed as
\begin{align}
\label{steering_vector_d}
    \bd &=
        \begin{pmatrix}
          \begin{bmatrix}
           (N-1)d_r \\
           (N-2)d_r\\
           \vdots \\
            0\\
          \end{bmatrix} \bigotimes
            \begin{bmatrix}
           0 \\
           d_t\\
           \vdots \\
            (M-1)d_t\\
          \end{bmatrix}
    \end{pmatrix},
  \end{align}
where $\bigotimes$ stands for the Kronecker product. Since $\tau_{mnl}$ is very small, hence, we ignore the term proportional to $\tau_{mnl}^2$ in (\ref{beat_signal_CDM}) and consider the beat signal as,
\begin{eqnarray}
\label{beat_signal}
s^{(l)}_{mni}(t) = \sigma_l \; e^{\displaystyle  - j \left(2 \pi f_c \tau_{mnl} + 2 \pi \beta t \tau_{mnl} - \varphi_i(m)\right)}.
\end{eqnarray}
The next step is the decoding process. To accomplish this goal, the signal at the output of each receiver will be multiplied by a specific code that has already been used for each transmitter and the result is expressed as
\begin{eqnarray}
\label{decoding}
\bar{\bs}^{(l)}_{mn} = \sum\limits_{i=1}^{N_c}\left(\hat{\bs}^{(l)}_{mni} \bigodot e^{j \displaystyle \varphi_i(m)}\b1_{1 \times \tilde N}\right).
\end{eqnarray}
In (\ref{decoding}), $\hat{\bs}^{(l)}_{mni} = [s^{(l)}_{mni}(1), s^{(l)}_{mni}(2), \cdots , s^{(l)}_{mni}(\tilde N)]$, where $\tilde N = T/t_s$, in which $t_s$ is the sampling time, and $\b1=[1,1, \cdots,1]_{(1 \times \tilde N)}$. The symbol $\bigodot$ stands for the Hadamard product.

In the case of having $L$ targets in front of the radar, the signal received by the radar is a linear combination of all these signals. Therefore, the signal received by the radar from all the signals $\bar{\bs}^{(l)}_{mn}$, for $l\in \{1,2,\cdots,L\}$, can be described as
\begin{eqnarray}
\label{total_signal}
\tilde{\tilde{\bs}}_{mn} =  \sum\limits_{l=1}^{L} \bar{\bs}^{(l)}_{mn}.
\end{eqnarray}
In the next section, we address the calibration process.
\section{Calibration}\label{calibration}
Calibration plays an important role in MIMO radar imaging. In array processing, measurement errors arise due to sensor gain and phase mismatches.

There are various reasons for these mismatches such as path length imperfections and different gain and phase per different elements. The major player is the phase mismatch which if it is not compensated for, the final image will be blurred and unfocused.
In this section, we present a very effective way to calibrate the FMCW MIMO radar.

To calibrate the system, we use a target at a specific range and incidence angle which we refer to them as $r_{\rm ref} \; \rm and \; \theta_{\rm ref}$, respectively. We then multiply the signals received from the targets with the reference signal and the result is given as
\begin{eqnarray}
\label{calib_1}
\underline{\bs}_{mn} = \tilde{\tilde{\bs}}_{mn}\bigodot \mathfrak{{\bar{s}}}^*_{mn},
\end{eqnarray}
where $\mathfrak{{\bar{s}}}^*_{mn}$ is a signal given as in (\ref{decoding}) for a target located at $(r_{\rm ref},\theta_{\rm ref} )$ and $^{(*)}$ stands for the complex conjugate operator.

The next step is to multiply (\ref{calib_1}) by the following term in order to restore the correct range of the targets
\begin{eqnarray}
\label{calib}
\tilde{\bs}_{mn} = \underline{\bs}_{mn} \bigodot \mathbf{\Im},
\end{eqnarray}
where $\mathbf{\Im} = [ 1 \; e^{j 4 \pi (f_c + (\frac{b}{\tilde{N}}))\frac{r_{ref}}{c}} \; \cdots \;  e^{j 4 \pi (f_c + (\frac{b}{\tilde{N}})(\tilde{N}-1))\frac{r_{ref}}{c}}]$.
In the next section, we present an image reconstruction technique to find the range and angle of arrival of the targets.
\section{Image Reconstruction}\label{image reconstruction}
In this section, we develop an algorithm based on the MUSIC method to reconstruct the range-angle information from the raw data and create a $\rm 2D$ image. To set the stage, we calculate the $\rm MN\tilde{N} \times MN\tilde{N}$ covariance matrix as follows
\begin{eqnarray}
\label{covariance_matrix}
 \bR = \mathbb{E} \{\underline{\bs}^T\underline{\bs}\},
\end{eqnarray}
where $\underline{\bs} = [\tilde{\bs}_{11} \; \tilde{\bs}_{12} \; \cdots \; \tilde{\bs}_{1N} \; \tilde{\bs}_{21} \;   \cdots \; \tilde{\bs}_{2N} \;  \cdots \;  \tilde{\bs}_{MN}]$ and $\tilde{\bs}_{mn} \; \rm for \; m \in \{ 1,2,\cdots,M\},\; n \in \{ 1,2,\cdots,N\}$ is given as in (\ref{calib}).

In practice, however, we only have access to a limited number of different realizations of $\bs$. Therefore, we can only obtain an estimate of (\ref{covariance_matrix}) which is called the sample covariance matrix.
The steering vector is defined as
\begin{eqnarray}
\label{stering_vector}
\ba(r,\theta) = e^{\displaystyle   j \left(2 \pi \frac{2}{c}(r + 0.5 \bd sin(\theta)) \bigotimes  \bg\right)},
\end{eqnarray}
where $\bg = [f_c \; f_c + \frac{b}{\tilde{N}} \; \cdots  \; f_c + \frac{b}{\tilde{N}}(\tilde{N}-1)]^T$.
Consequently, the image provided by the  MUSIC algorithm for a target located at $r$ with angle of arrival $\theta$ is expressed as
 \begin{align}
\label{music}
\mathcal{I}_{\rm{MUSIC}}(r,\theta) =  {\frac{{{\ba^{\dagger}(r,\theta)}\ba(r,\theta)}}{{\ba^{\dagger}(r,\theta)}{\bE}{\bE^{\dagger}}\ba(r,\theta)}},
\end{align}
where ${\ba}(r,\theta)$ is the steering vector given in (\ref{stering_vector}) and the columns of $\rm {\bE} \in \mathbb{C}^{MN\tilde{N}\times (MN\tilde{N}-L)}$ are the eigenvectors of the sample covariance matrix corresponding to the smallest $\rm MN\tilde{N}-L$ eigenvalues with $\rm L$ being the effective dimension of the signal subspace. Finally, $^{(\dagger)}$ stands for the complex conjugate transpose operator.
\section{Experimental Results}\label{result}
\begin{figure}
\centering
\includegraphics[height=7cm,width=8cm]{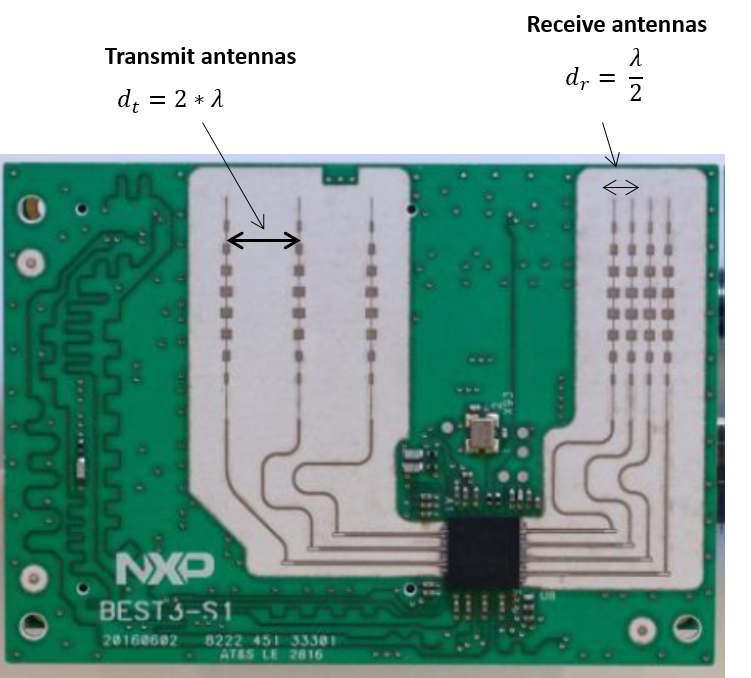}
\caption{The sparse MIMO FMCW radar from NXP semiconductors with its transmit and receive antennas as well as the element spacing.
\label{fig:radar}}
\end{figure}

In this section, we present the experimental results. We have used a FMCW radar from NXP semiconductors operating at $\rm 79\;GHz$ with $3$ transmit and 4 receive antennas. The radar chip, along with its transmit and receive antennas, has been shown in Fig.~\ref{fig:radar}.

The bandwidth and the length of the transmitted up-chirp signal are $\rm 1.5\;GHz$ and $\rm 25.6\;\mu s$, respectively. The output power is $\rm 5\;dBm$. The gain for both the transmit and the receive antennas is $\rm 14\;dB$ at boresight. The sampling frequency of the ADC is $\rm 20\;MHz$.

We have set the maximum range at $\rm 3 \; m$ and based on the values given for the bandwidth and the chirp time, the beat frequency for a target at $\rm R = 3\;m$ is calculated as $ f_b = \frac{b}{T}\frac{2R}{c} = 1.17 \; {\rm MHz}$. Therefore, we have downsampled the signals by factor $8$.

From Fig.~\ref{fig:radar}, we see that $d_r = \frac{\lambda}{2}$ and $d_r = 2\lambda$ and as a result, $d_t = 4d_r$ which is an indication that virtual MIMO processing is possible and we can effectively increase the number of receive elements from 4 to $3\times 4$.

To implement the idea of CDM-based virtual FMCW MIMO radar, we have used Walsh code \cite{Walsh} of length 8. We utilize the following three codes for the three transmitters \\
$TX1 = [ +1\; +1\; +1\; +1\; -1\; -1\; -1\; -1]$,\\
$TX2 = [ +1\; +1\; -1\; -1\; +1\; +1\; -1\; -1]$,\\
$TX3 = [ +1\; -1\; +1\; -1\; +1\; -1\; +1\; -1]$,\\
where +1 and -1 represent 0 and 180 degree phase shift, respectively.

\begin{figure}
\centering
\includegraphics[height=6cm,width=8cm]{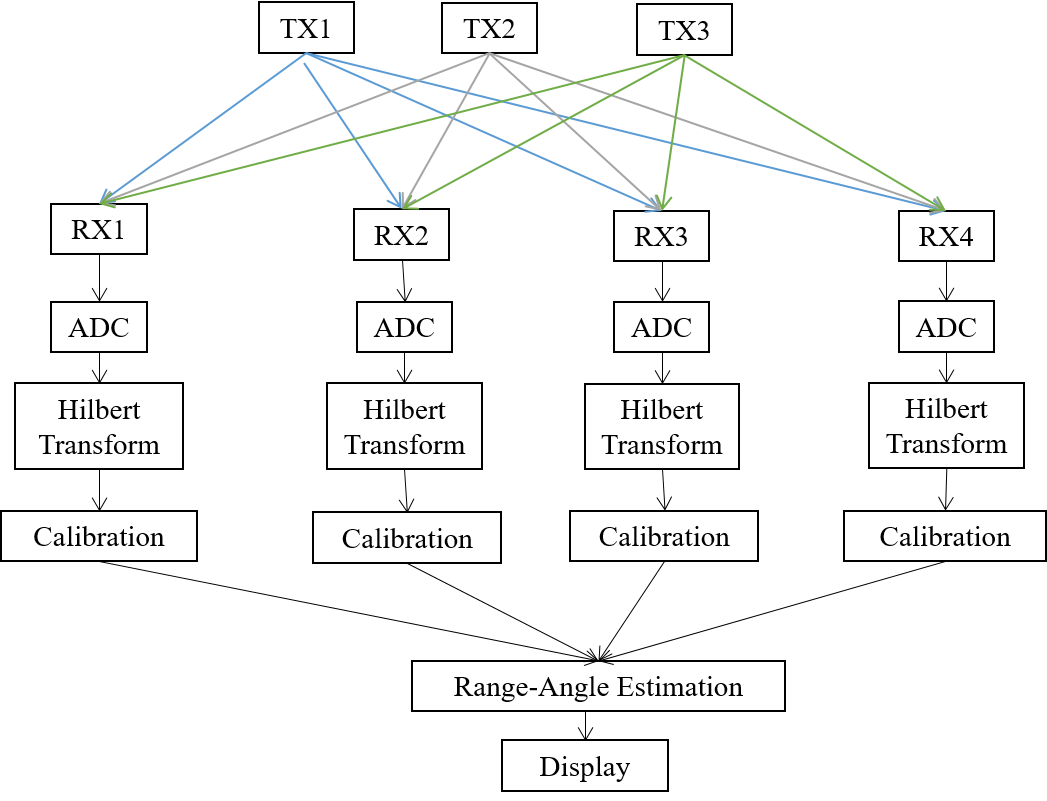}
\caption{The flowchart of the algorithm.
\label{fig:Flow_chart}}
\end{figure}

\begin{figure}
\centering
\includegraphics[height=5cm,width=8cm]{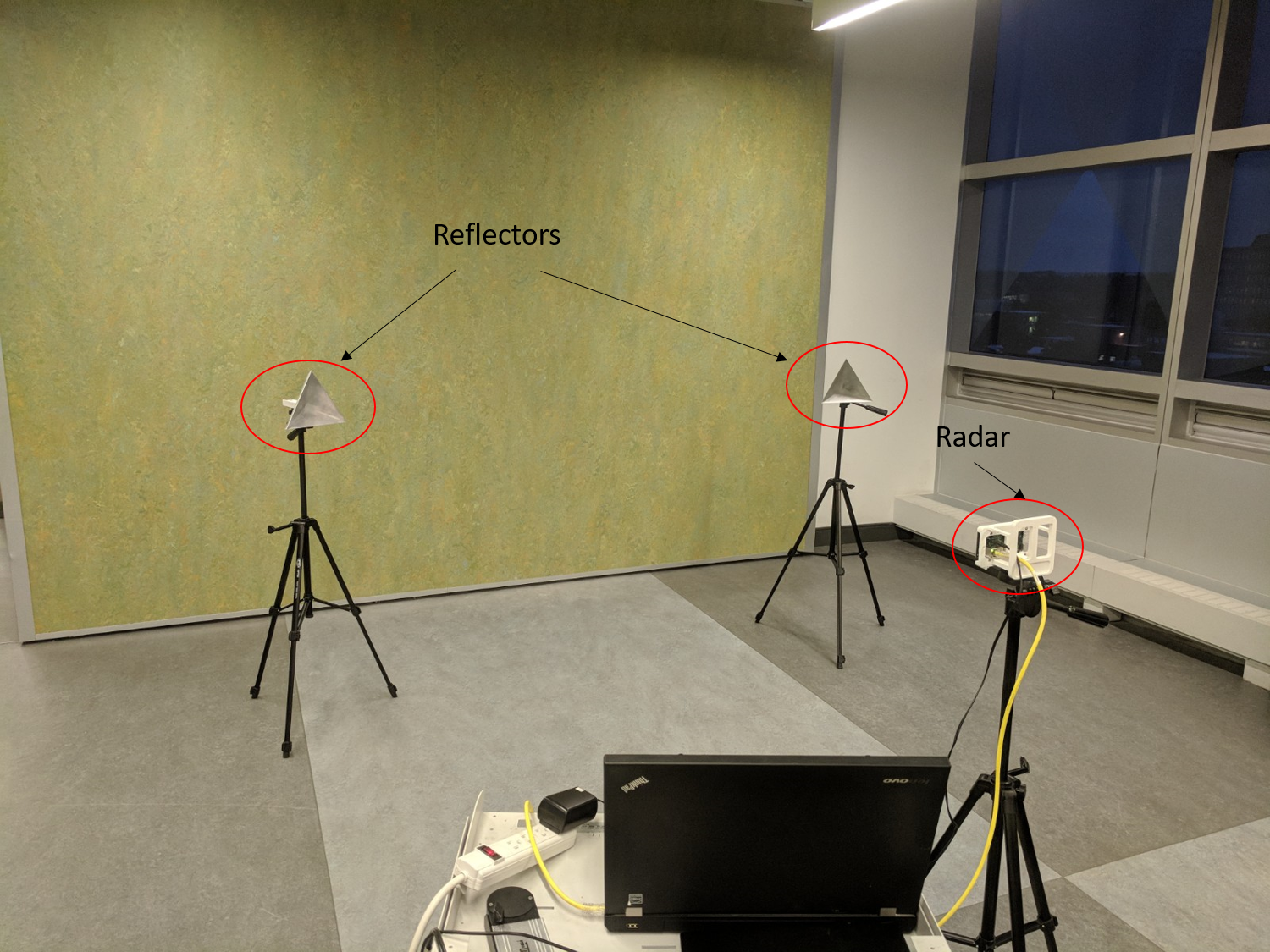}
\caption{The experimental set-up for two $\rm 20 \; dBsm$ reflectors in front of the radar.
\label{fig:exp_set_up}}
\end{figure}

\begin{figure}
\centering
\psfrag{Azimuth Angle(degree)}{\tiny \rm Azimuth Angle(degree)}
\psfrag{Range(m)}{\tiny \rm Range(m)}
\includegraphics[height=7cm,width=9cm]{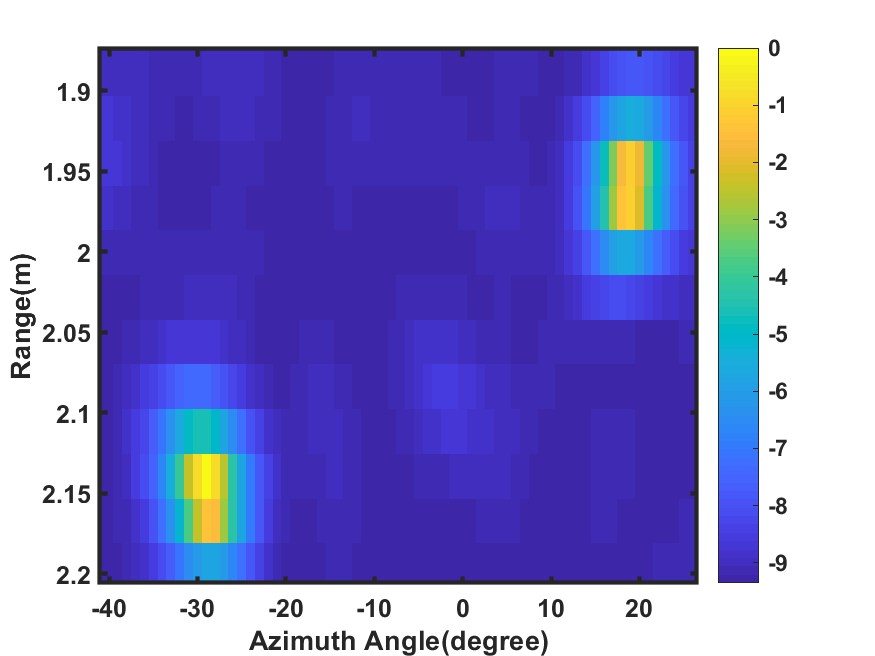}
\caption{The result of applying (\ref{music}) to the experimental data gathered from the set-up shown in Fig. ~\ref{fig:exp_set_up}. The colorbar is in dB unit.
\label{fig:cdm_image}}
\end{figure}

Fig.~\ref{fig:Flow_chart} shows the flowchart of the proposed algorithm for our CDM-based virtual FMCW MIMO radar. The purpose of the Hilbert transform \cite{Hilbert} block is to transform the output data of the ADC from real numbers to complex numbers and prepare the data for calibration.
Fig.~\ref{fig:exp_set_up} shows the experimental set-up where there are two $20\;\rm dBsm$ reflectors in front of the radar at $(195\;\rm cm,19^{\circ})$, $(214\;\rm cm,-29^{\circ})$.
Fig.~\ref{fig:cdm_image} illustrates the result of applying (\ref{music}) to the experimental data. The dimension for the signal subspace has been set to $2$. We have used $42$ different realizations of the received signal to obtain an estimate of the covariance matrix given in (\ref{covariance_matrix}).
\section{Conclusions}
We presented a full description of $2 \rm D$ radar imaging process for a CDM-based virtual FMCW MIMO radar. We started with the problem formulation and then addressed the system calibration at the baseband level. Based on the presented model, we described a $2 \rm D$ MUSIC algorithm for image reconstruction. Finally, we presented the result of applying the MUSIC technique to the data gathered from a CDM-based virtual FMCW MIMO radar operating at $79\;\rm GHz$ with $3$ transmit and $4$ receive antennas using Walsh code of length $8$.

Radar imaging based on coded signals, which was described in this paper, can easily address the interference problem in crowded environments where multiple FMCW radars are operating in close proximity of each other.
\section{Acknowledgement}
The authors are grateful to NSERC, NXP Semiconductors and MAGNA for their supports.

\bibliographystyle{IEEEtran}
\bibliography{Biblio}

\end{document}